\documentclass[12pt]{article} 
\usepackage{graphicx}
\usepackage{amsmath,amssymb}

\begin{document}
\baselineskip 21pt
\setcounter{totalnumber}{8}

\bigskip

\centerline{\Large \bf Extended Gaseous Disk }
\centerline{\Large \bf in the S0 galaxy NGC 4143}

\bigskip

\centerline{\large O.K. Sil'chenko$^1$, A.V. Moiseev$^{2,1}$, and D.V. Oparin$^2$}

\noindent
{\it Sternberg Astronomical Institute of the Lomonosov Moscow State University, Moscow, Russia}$^1$

\noindent
{\it Special Astrophysical Observatory of the Russian Academy of Sciences, Nizhnij Arkhyz, Russia}$^2$

\vspace{2mm}
\sloppypar 
\vspace{2mm}

\bigskip

{\small 
\noindent
We present our results of the spectroscopic study of the lenticular galaxy NGC 4143 -- an
outskirt member of the Ursa Major cluster. Using the observations at the 6-m SAO RAS telescope
with the SCORPIO-2 spectrograph and also the archive data of panoramic spectroscopy with the SAURON
IFU at the WHT, we have detected an extended inclined gaseous disk which is traced up to a distance 
of about 3.5 kpc from the center, with a spin approximately opposite to the spin of the stellar disk. 
The galaxy images in the H-alpha and [NII]$\lambda$6583 emission lines obtained at the 2.5-m CMO
SAI MSU telescope with the MaNGaL instrument have shown that the emission lines are excited by a
shock wave. A spiral structure that is absent in the stellar disk of the galaxy is clearly seen in the brightness
distribution of ionized-gas lines (H-alpha and [NII] from the MaNGaL data and [OIII] from the SAURON
data). A complex analysis of both the Lick index distribution along the radius and of the integrated
colors, including the ultraviolet measurements with the GALEX space telescope and the near-infrared
measurements with the WISE space telescope, has shown that there has been no star formation in the
galaxy, perhaps, for the last 10 Gyr. Thus, the recent external-gas accretion detected in NGC 4143 
from its kinematics, was not accompanied by star formation, probably, due to an inclined direction of 
the gas inflow onto the disk.
}

\clearpage

\section{INTRODUCTION}

Lenticular galaxies are traditionally assigned to morphological early type, with their red color and an absence
of visible traces of current star formation. However, many surveys of representative samples of nearby early-type galaxies, 
for example, search for cold gas within the ATLAS-3D project (Cappellari et al. 2011), show that very many and, in sparse 
environments, most of the lenticular galaxies possess extended gaseous disks (Welch and Sage 2003; Sage and Welch 2006; 
Welch et al. 2010; Davis et al. 2011).
However, only less than half of the gas-rich S0 galaxies exhibit at least faint signs of current star formation
(Pogge and Eskridge 1993). If the field lenticular galaxies accrete an external cold gas just as the spiral
ones, why does star formation not proceed properly in their disks? In attempting to answer this question,
we have recently analyzed gas velocity fields for a small sample of nearby S0 galaxies in which extended,
regularly rotating gaseous disks are observed (Sil'chenko et al. 2019). It turned out that an inclined
direction of the gas infall onto the galactic disk could be responsible for the absence of star formation: in
this case, the gas experiences shock excitation, heats up, and cannot collapse into stars. Thus, not only
the very presence of a gas, but also the direction from which it comes into the galaxy does matter for
the shaping of its morphological type. 

In this paper we present one more lenticular galaxy
in which there is an extended gaseous disk, but, apparently, there are no young stars. This is
the moderately luminous lenticular galaxy NGC~4143 ($M_H =-23.4$, NED). Its gaseous disk detected
by us consists of an ionized warm gas; this time no cold neutral hydrogen and molecular gas have
been reported (Young et al. 2011; Serra et al. 2012). The galaxy belongs to the Ursa Major cluster (Tully
et al. 1996), in which there are many spiral galaxies; the cluster has not yet relaxed structurally (it consists of
several rich groups, Karachentsev et al. 2013; Pak et al. 2014), and unlike most galaxy clusters, it does not reveal any 
signs of a hot intergalactic medium affecting the galaxies (Verheijen and Sancisi 2001). Few
lenticular galaxies, Ursa Major members, actively accrete intergalactic neutral hydrogen (NGC~4026
and NGC~4111 by Serra et al. 2012, NGC~4138 by Jore et al. 1996). The galaxy NGC~4143 itself is located
on the cluster periphery and, specifically, nothing has been noticed around it in the 21-cm line. However,
the [OIII]$\lambda$5007 emission line was found in the galaxy within the ATLAS-3D spectroscopic survey
(Cappellari et al. 2011), and the ionized-gas velocity field at the galactic center turned out to be quite
unusual. We decided to undertake our own study of this lenticular galaxy, which has well fitted into the
problematics of the absence of star formation in the gaseous disks of lenticular galaxies.

\section{OBSERVATIONS AND DATA ANALYSIS}

Our long-slit spectroscopy was taken on the night of March 2/3, 2016, with the SCORPIO-2 focal reducer 
at the 6-m SAO RAS telescope (Afanasiev \& Moiseev 2011) with the VPHG1200@540 grism and a spectral resolution 
of 5~\AA. A slit of 1 arcsec in width and about $6^{\prime}$ in length was aligned with the major axis of the 
galactic isophotes at $PA=144$~deg. The total exposure was 1 hour at a seeing of about 2.5 arcsec. We measured 
the line-of-sight velocities of the stellar component by cross-correlating pixel-by-pixel spectra taken along 
the slit at various distances from the galactic center, with the spectrum of the K1.5 III star HD 72184 taken 
on the same night with the same set up. The data have turned out to be deep enough to measure the stellar 
kinematics up to the optical boundaries of the galaxy. The line-of-sight velocities of the gaseous component 
were measured through Gaussian analysis of the blend of the H$\alpha + \mbox{[NII]}\lambda$6548,6583 emission 
lines with the H$\alpha$ absorption line. The results of our kinematic measurements, namely, the radial profiles 
of the line-of-sight velocities of the gas and stars, are presented in Fig.~1.

\begin{figure*}[!h]
\includegraphics[scale=0.8]{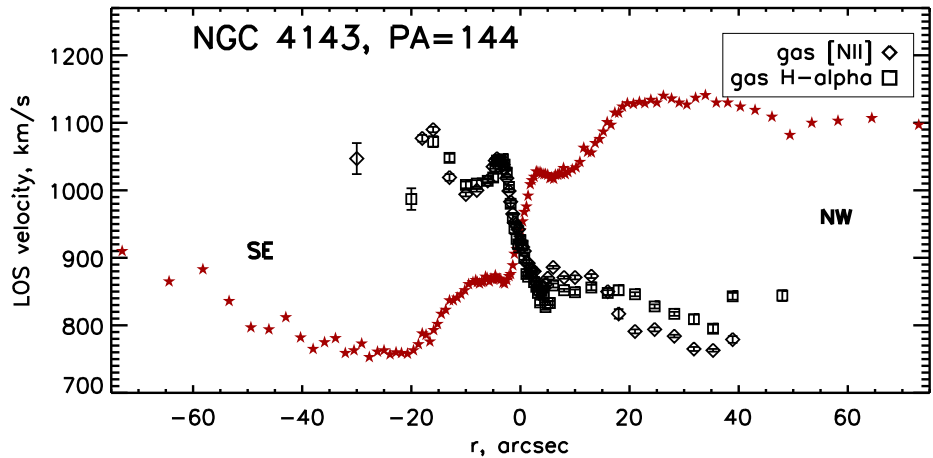}
\caption{
Line-of-sight velocity profiles for the stellar component (red stars) and for the ionized gas
(black signs for various emission lines).}
\label{longslit}
\end{figure*}

As a supplement to the extended one-dimensional kinematic cuts obtained with a long slit, we had calculated
two-dimensional line-of-sight velocity maps of the gas and stars for the galactic center from the IFU datacube
obtained with the SAURON integral-field spectrograph (Bacon et al. 2001). The galaxy NGC 4143 was observed as
a part of the ATLAS-3D project (Cappellari et al. 2011) with the 4.2-m William Herschel Telescope (WHT). 
The raw data were retrieved by us from the ING (Isaac Newton Group) open archive of the Cambridge Institute 
of Astronomy and were reduced through our original technique (Sil'chenko 2005). The field of view of the SAURON
spectrograph is $33^{\prime \prime} \times 41^{\prime \prime}$, one spatial element is $0.94 \times 0.94$ arcsec, 
the spectral range explored is 4800--5400~\AA, and the spectral resolution is about 4~\AA. The stellar and gas
velocity fields for NGC~4143 are shown in Fig.~2. They were analyzed by the tilted-ring method in Moiseev's
modification (the DETKA code; Moiseev et al. 2004). The orientation of the kinematic major axis of both
components, stellar and gaseous, was traced. The kinematic major axis should coincide with the line
of nodes of the disk in the case of circular rotation. Based on the SAURON data, we managed to extend
our measurements of the orientation of the kinematic major axis up to a distance of about $20^{\prime \prime}$ 
from the center.

\begin{figure*}[p]
\begin{tabular}{c c}
\includegraphics[width=8cm]{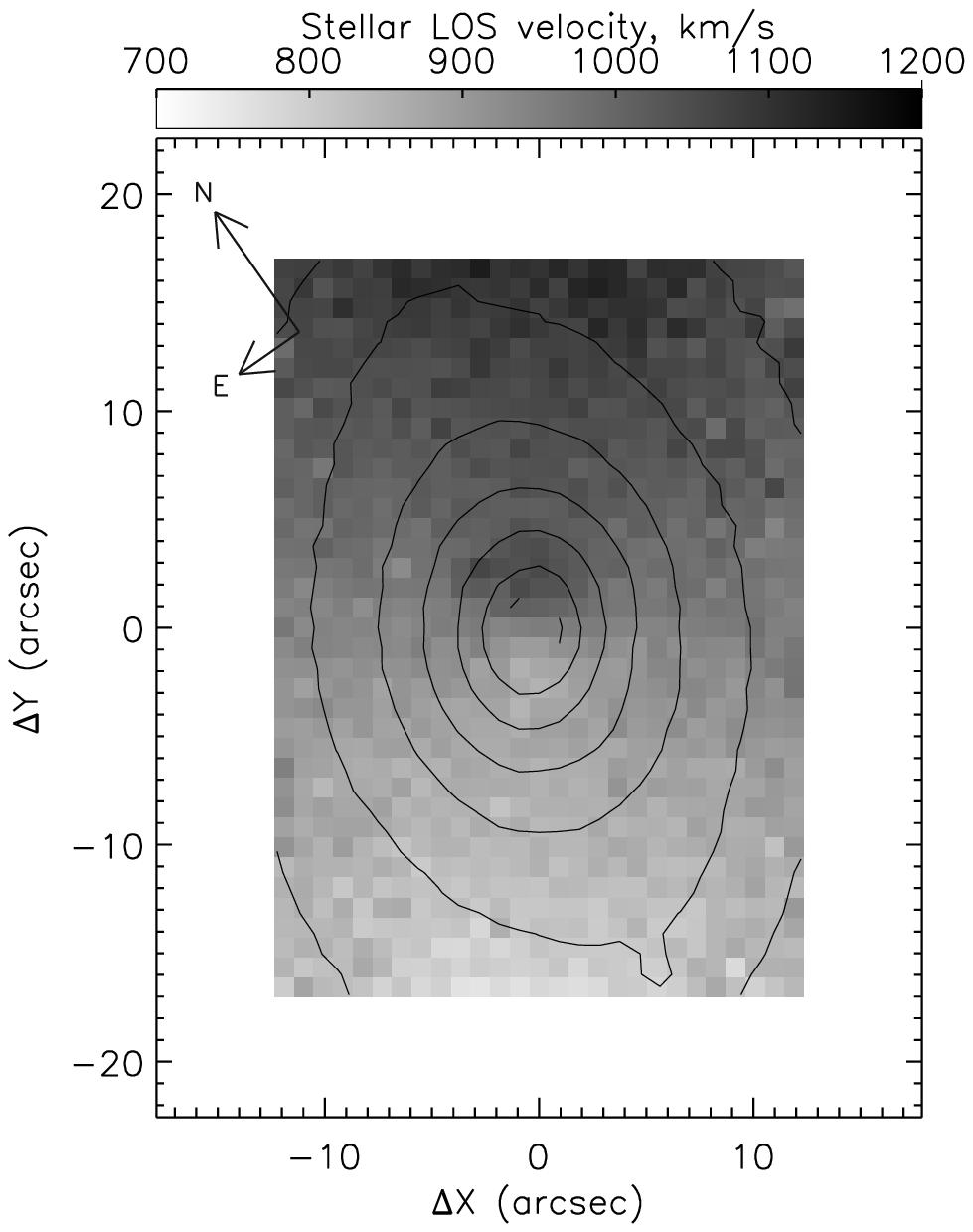} &
\includegraphics[width=8cm]{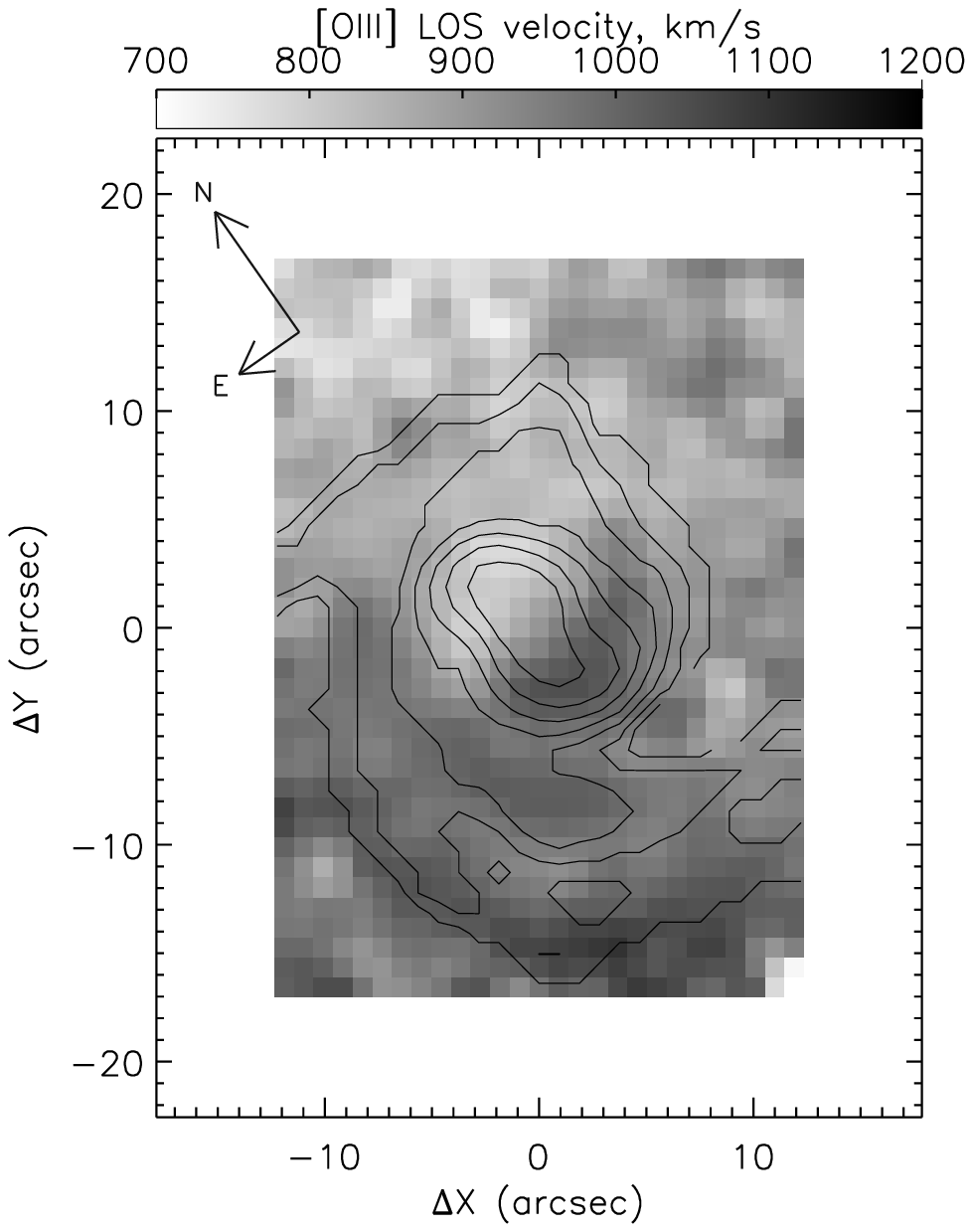}\\
\end{tabular}
\caption{
IFU SAURON line-of-sight velocity fields for the stellar component ({\it left}) and for the ionized gas
({\it right}) in the center of NGC~4143; the gas velocities are calculated from the [OIII]$\lambda$5007
spectral positions. The map orientation is designed by upper left arrows pointing to the north and to the east.
The isophotes overposed show the 5100~\AA\ continuum distribution at the left plot and [OIII] emission-line
flux distribution at the right plot.}
\label{n4143sau}
\end{figure*}

A photometric analysis of the galactic structure has been presented in the literature more than once. A
two-dimensional decomposition of the NGC~4143 image was undertaken by Laurikainen et al. (2010, 2011) and 
P.~Erwin in a series of papers (Erwin and Sparke 2003; Erwin et al. 2005, 2008). We additionally
performed an isophotal analysis of the g- and r-band images for NGC~4143 based on the SDSS data,
release 9 (Ahn et al. 2012), which results are shown in Fig.~3.

\begin{figure*}[p]
\begin{tabular}{c c}
\includegraphics[width=8cm]{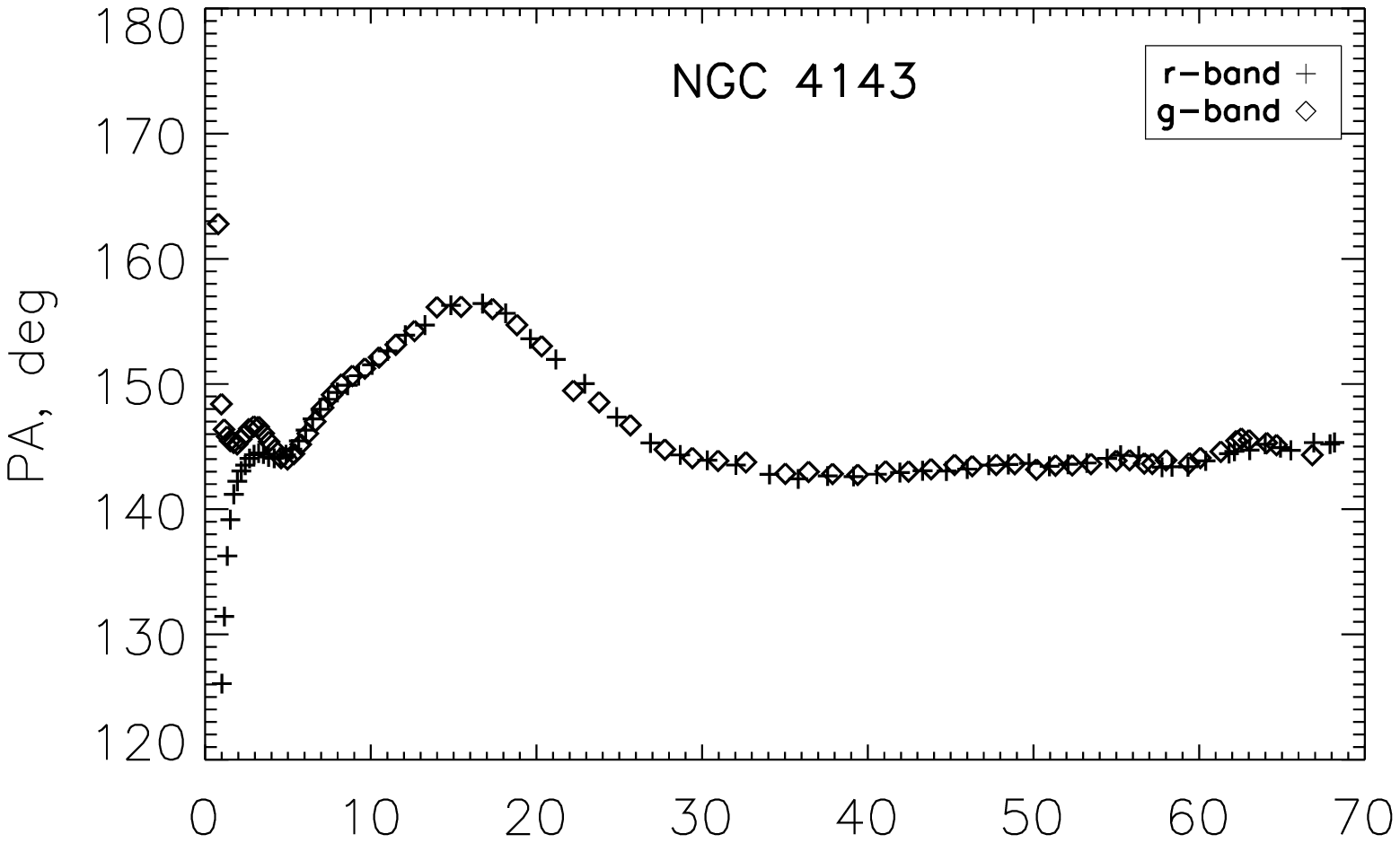} &
\includegraphics[width=8cm]{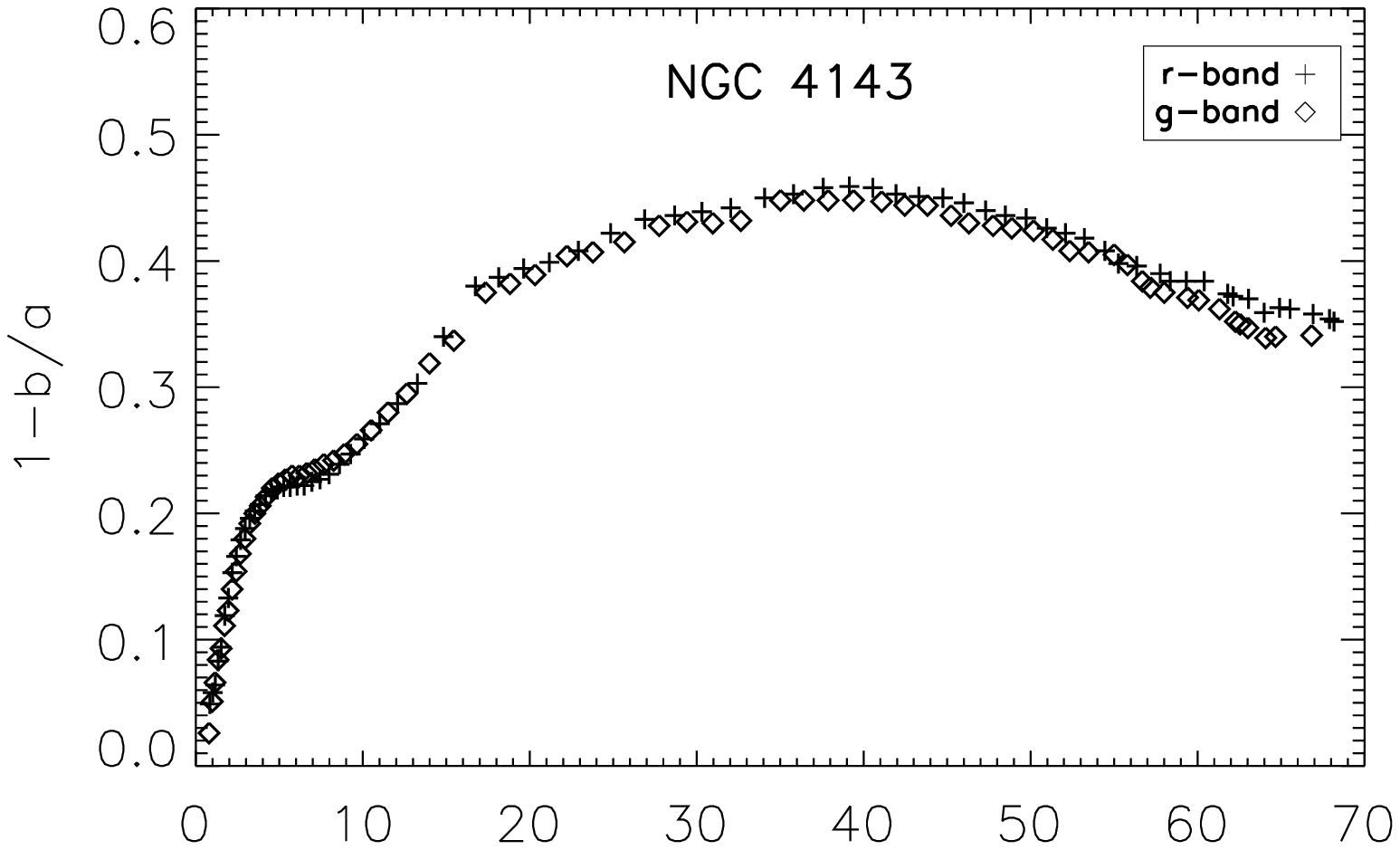}\\
\end{tabular}
\caption{Isophote analysis results derived with the SDSS images: radial profiles of the major axis
position angle ({\it left})and isophote ellipticity ({\it right}).
}
\label{n4143sdss}
\end{figure*}

We also carried out observations at the 2.5-m CMO SAI MSU telescope (Kornilov et al. 2014)
with a new instrument -- MaNGaL (Mapper of Narrow Galaxy Lines, Moiseev et al. 2020). MaNGaL
is a tunable-filter photometer based on a scanning Fabry–Perot interferometer with an instrumental
profile width (FWHM) $\sim 15$~\AA. The detector, a low-noise iKon-M934 $1024 \times 1024$-pixel CCD camera,
was used in a $2 \times 2$ binning mode to save the readout time and to reduce the noise. The final scale
was $0.66^{\prime \prime}$ per pixel. During the observations we subsequently accumulated the frames when 
the filter passband was centered onto the redshifted H$\alpha$ and [NII]$\lambda$6583 emission lines
(given the mean velocity of the galaxy and the heliocentric correction) and to the continuum shifted
by 50~\AA\ from H$\alpha$ blueward. Such series of exposures allow to eliminate the effects
of atmospheric transparency and seeing variations. The observations were performed on the night of April 13/14, 2018,
with total exposures of 2400 s in H$\alpha$ and continuum and 2100 s in [NII]$\lambda$6583; the spatial resolution of
the combined images is $2.3^{\prime \prime}$. The reduction of the MaNGaL data is similar to that of ordinary direct
images with narrow filters and is described in Moiseev et al.(2020). After the continuum subtraction, we
obtained maps of the total galaxy field in the H$\alpha$ and [NII]$\lambda$6583 emission lines. This allowed us not
only to study the ionized-gas morphology, but also to estimate the ratios of the strong nitrogen and hydrogen
emission lines over the entire galactic disk by dividing one two-dimensional emission line intensity
distribution by the other; the ionized-gas excitation mechanism can be constrained using the measurements of this ratio.
Since the relative intensity of the emission lines is low -- in the galactic disk EW(H$\alpha$) falls within the
range $0.5-1.5$~\AA, -- we checked the continuum subtraction accuracy in the MaNGaL data using our
spectroscopic measurements with the SCORPIO-2. We chose the normalization by which the continuum images
were multiplied before subtraction to reach the best agreement between the observed distributions of the equivalent
widths EW(H$\alpha$) and EW([NII]) along the major axis of the galaxy from our long-slit spectroscopy and in the images
obtained with MaNGaL. The difference between this normalization and that determined from background stars
(as the standard technique suggests when working with narrow-band images) has been found to be only 2\% --4\%, which is
within the reasonable assumptions about the difference between the averaged spectral energy distributions
of the background stars and of the galaxy itself.

\section{RESULTS OF OUR MEASUREMENTS}

Both the line-of-sight velocity profiles of the gas and stars along the major axis (Fig.~1) and the two-dimensional
line-of-sight velocity maps of the gas and stars for the central part of NGC~4143 (Fig.~2)
show that the gas in the galaxy rotates in the opposite direction with respect to the stars; our long-slit
observations demonstrate this counter-rotation over the full extension of the gaseous disk, up to $45^{\prime \prime}$
(3.5~kpc) from the center. If we look at Fig.~2b, where the two-dimensional gas velocity field in the central part 
of the galaxy measured through the [OIII]$\lambda$5007 emission line is presented, then we will see noncircular motions: 
the surface brightness distribution of the [OIII]$\lambda$5007 emission line looks like a one-armed spiral along which
an excess of the line-of-sight velocity of the ionized, highly excited gas is observed. The switch of the
relative gas velocities from positive to negative ones when passing across this spiral arm
can be a manifestation of the radial gas inflow toward the center along the spiral that is a shock front.

\begin{figure*}[!h]
\includegraphics[width=0.8\hsize]{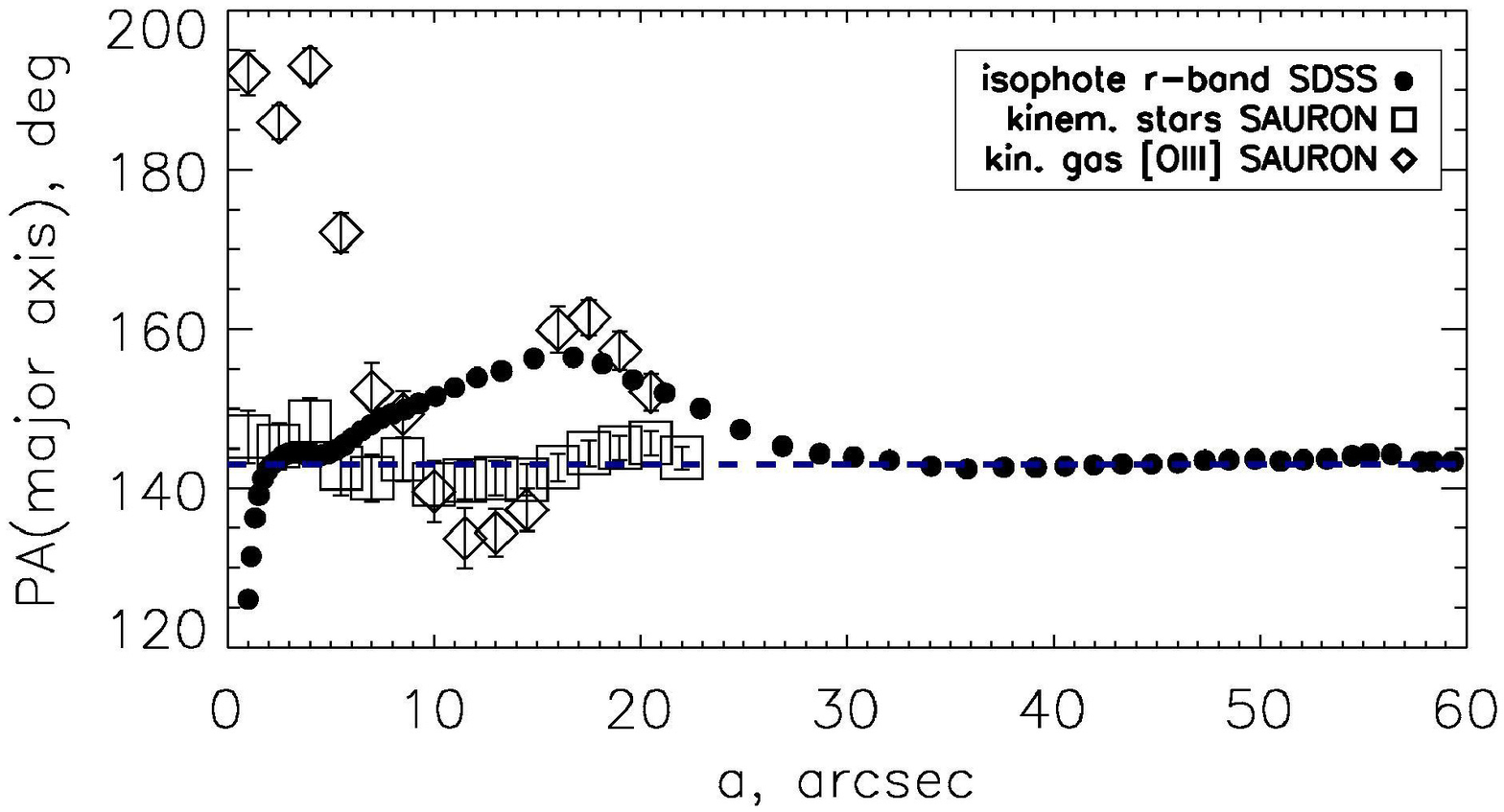}
\caption{The comparison of the major axis orientations, both photometric and kinematic, for the stars
and ionized gas. We have added 180 deg to the position angle of the stellar kinematical major axis.
The horizontal dashed line marks the orientation of the stellar-disk line of nodes determined from the
outermost isophotes.}
\label{pacomp}
\end{figure*}

Figure~4 compares the kinematic and photometric major-axis orientations derived by the tilted-ring method from
the two-dimensional velocity fields. In the case of circular rotation of a round disk within an axisymmetric potential, 
the major axis of elliptical isophotes (since a circle at an arbitrary inclination to our line of sight should
appear in projection precisely as an ellipse) will coincide with the line of nodes of the disk plane, while the
maximum rotation velocity projection onto the line of sight will also be precisely at the line of nodes. Hence,
the direction of the apparent maximum line-of-sight velocity gradient of the galaxy''s rotation
in this case should also be along the line of nodes: the photometric major axis should coincide with the
kinematic one. Otherwise, if the 'test point' rotates in a nonaxisymmetric potential, for example, at the
center of the disk of a barred galaxy, then, as Vauterin and Dejonghe (1997) showed in their simulations, the
major axes, photometric and kinematic, should turn in opposite sense with respect to the line of nodes (for a discussion
and references, see also Moiseev et al. 2004). What do we actually see in Fig.~4 for the 'barred' galaxy NGC 4143?
Our measurements of the orientation of the kinematic major axis of the stellar component lie strictly
along the line of nodes, despite the proposed presence of a bar claimed by Laurikainen et al. (2010, 2011) and Erwin
and Sparke (2003) based on their analysis of the photometric data. The orientation of the kinematic major
axis of the gaseous component not just deviates from the line of nodes of the galactic stellar disk, but it
coincides with the orientation of the isophotes at a radius $R=15^{\prime \prime} -20^{\prime \prime}$.
Whereas the turn of the kinematic major axis of the gaseous component at the very center of the galaxy, 
at $R < 5^{\prime \prime}$, can be explained as noncircular gas motions (net radial flows will
show a turn of the orientation of the maximum line-of-sight velocity gradient by 90 degrees relative to
the kinematic axis of circular rotation), in the region of the maximum twist of the isophotes the gas kinematic 
behavior cannot be interpreted as noncircular rotation. This more closely resembles an inclined disk that,
besides, also has a small stellar component. Indeed, if we look at the results of our isophotal analysis of the
galaxy images (Fig.~3), then we see that at the radius of the maximum turn of the isophotes,  $R=17^{\prime \prime}$, 
which other researchers marked as the bar end, the ellipticity of the isophotes is strictly equal to
the ellipticity of the outer disk isophotes. If the bar were responsible for the turn of the isophotes, then at
such a small deviation of the bar orientation from the line of nodes an increased ellipticity of the isophotes
could have been expected precisely at this radius, which is not observed. It seems that both isophotal
and kinematic analyses of the data for the central region of NGC~4143 more likely provide evidence for
a central inclined disk containing a stellar component and associated with the orientation of the plane of gas
rotation, namely, with the gaseous disk.

Using the MaNGaL instrument at the 2.5-m CMO SAI MSU telescope, we obtained images of the total galaxy field 
in the narrow [NII]$\lambda$6583 and H$\alpha$ emission lines and as well as a map of the ratios of the fluxes
in these emission lines by dividing one by the other (Fig.~5). The ratio of low-excitation lines, which the
[NII]$\lambda$6583 line is, to the hydrogen emission line is a good indicator of the gas excitation mechanism. 
The boundary value of the [NII]$\lambda$6583-to-H$\alpha$ ratio is 0.5: for the gas excited by young massive 
stars this ratio is always less than 0.5 (Kewley and Ellison 2008). The second necessary signature of gas excitation 
by young stars is a sufficiently large equivalent width of the H$\alpha$ emission line: although its lower limit 
in the presence of star formation is 1~\AA\ (Cid Fernandes et al. 2010), it is usually required that EW(H$\alpha$) 
exceeds 3~\AA. If we analyze Fig.~5 from the standpoint of these criteria, then we will find that the gas in NGC~4143
is excited not by young stars, but more likely by shock waves. The morphology of the galaxy image in the emission
lines shows the presence of spiral arms, with the spirals in the [NII] and H$\alpha$ images coinciding with
those in the [OIII] emission line from the SAURON data within the central region. The location of these spiral arms
closely traces the maximum ratio of the fluxes in the nitrogen and hydrogen emission lines:
[NII]/H$\alpha > 1.5$ (Fig.~5, lower right). Interestingly, if we rely on the spiral pattern of this last figure, 
then a minimum equivalent width of the H$\alpha$ emission line less than 1~\AA\ will correspond to a minimum ratio
[NII$\lambda$]6583/H$\alpha \sim 0.5$, by outlining this pattern. Thus, the gas in the regions with a minimum ratio
of the fluxes in the nitrogen and hydrogen emission lines is also excited not by young stars, but most
likely by a shock, because these spirals are spatially far from the LINER-type active galactic nucleus
in NGC~4143. Such excitation of the gas is absolutely consistent with its kinematics. Indeed, when the gaseous disk
rotates with a significant inclination with respect to the equatorial plane of the gravitational potential of
the stellar disk, during each crossing of the potential well of the stellar disk by gas clouds a shock wave
develops in the gas (Wakamatsu 1993). The one-armed spiral that we see in the surface brightness maps of 
the emission lines and that is absent in the surface brightness maps in continuum and broad-band
colors (i.e., it is associated neither with the stellar population nor with dust), can be a manifestation
of the two-stream instability due to the dynamical interaction of two counter-rotating subsystems at
the galactic center, stellar and gaseous ones (Lovelace et al. 1997).

\begin{figure*}[p]
\centerline{
\includegraphics[height=0.5\textwidth]{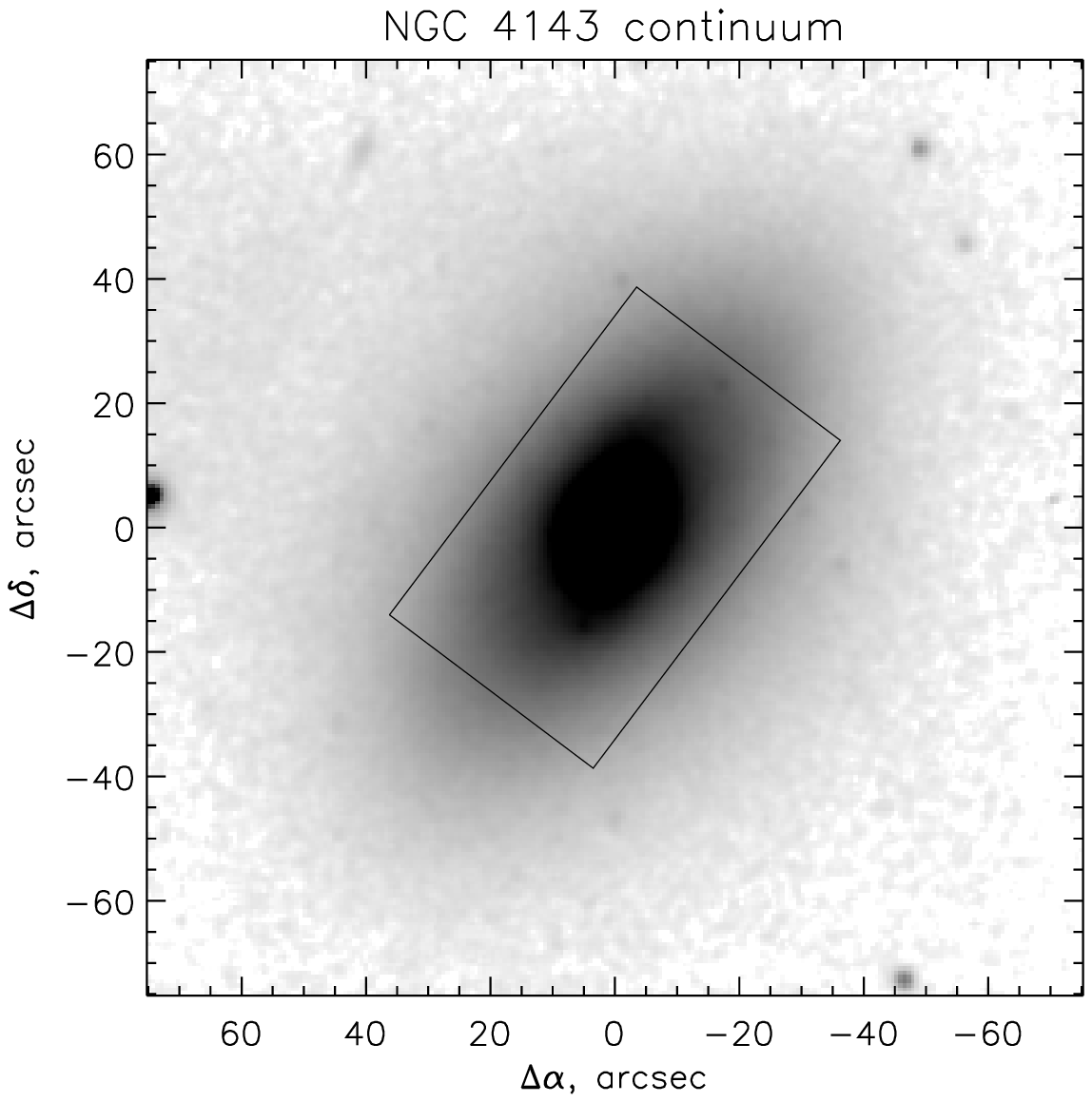}
\includegraphics[height=0.5\textwidth]{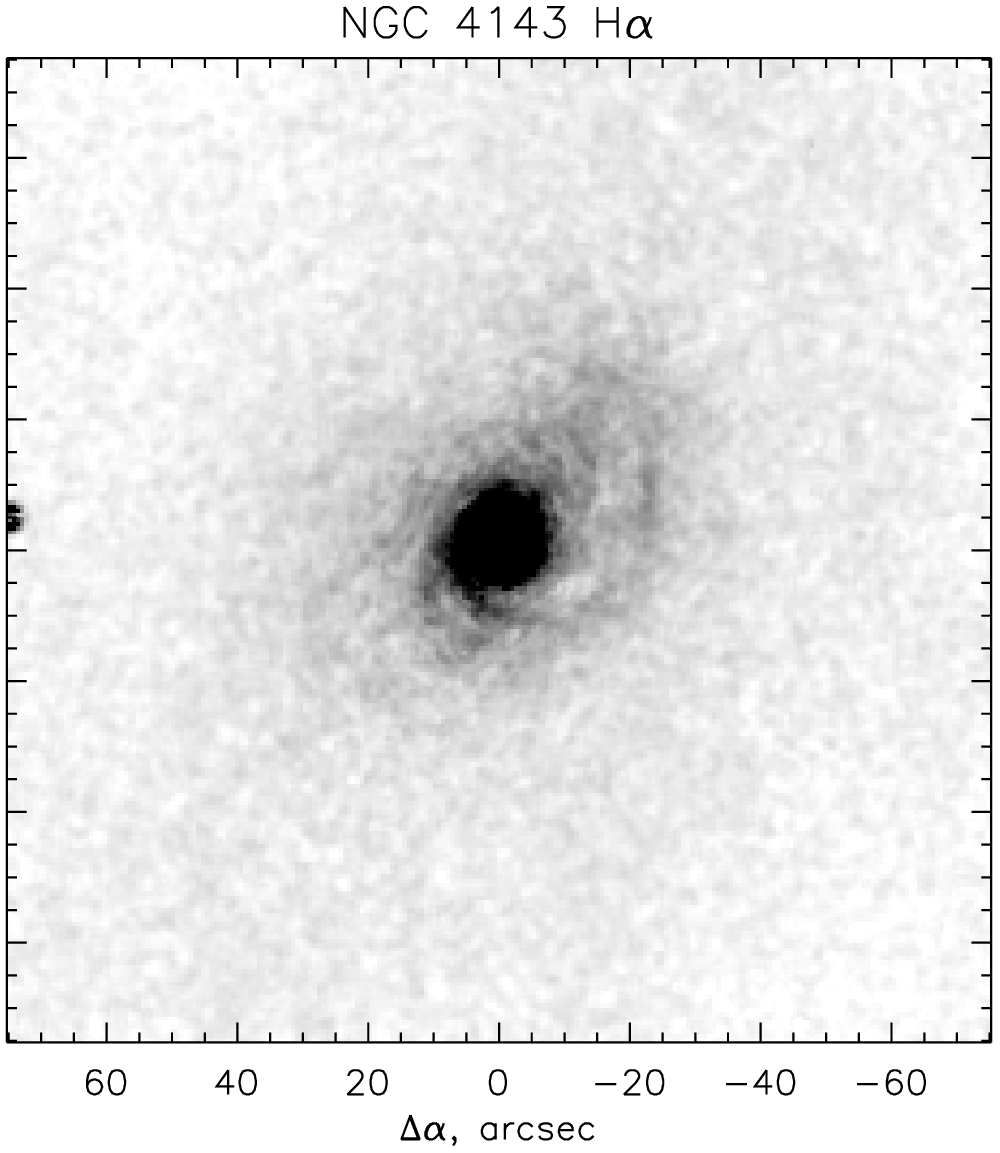}
}
\centerline{
\includegraphics[height=0.5\textwidth]{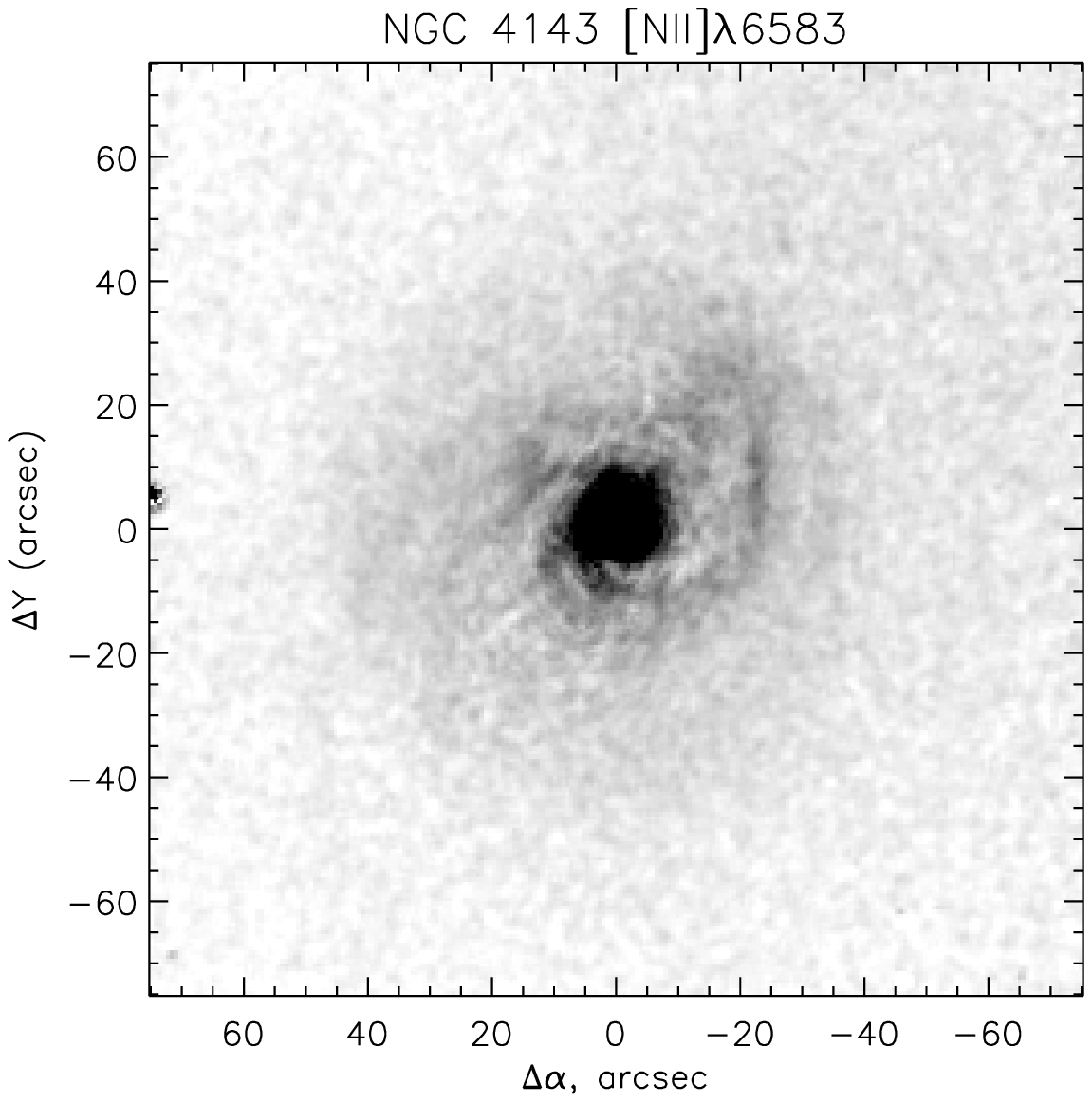}
\includegraphics[height=0.5\textwidth]{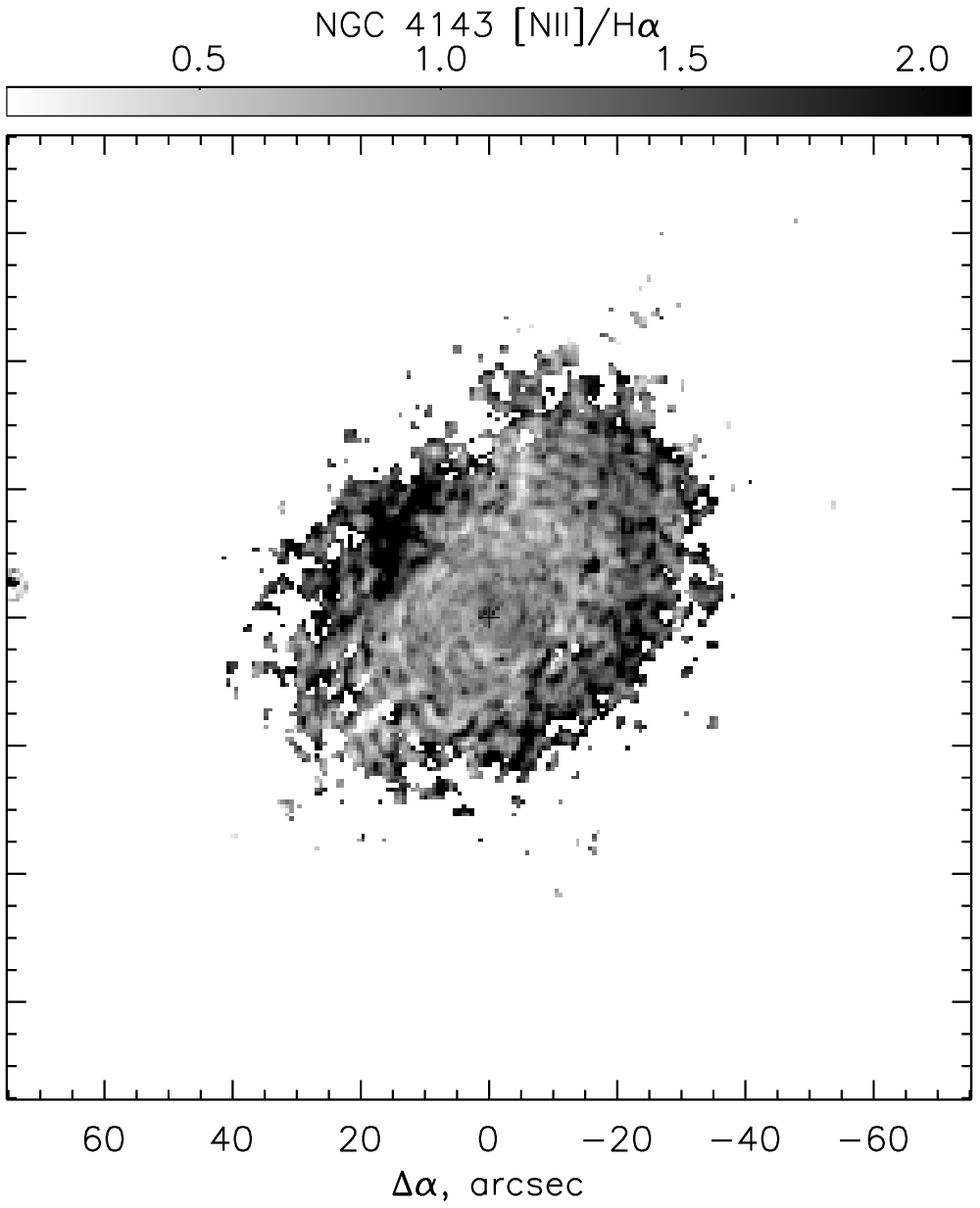}
}
\caption{The images of NGC~4143 in the narrow-band filters cutting the red continuum and redshifted emission lines
H$\alpha$ and [NII]$\lambda$6583 (the latters -- with the continuum subtracted), obtained with the instrument MaNGaL 
at the 2.5m telescope of the Caucasus Mountain Observatory of the Sternberg Astronomical Institute. The last
right-bottom plot presents the flux ratio of the emission lines H$\alpha$ and [NII]$\lambda$6583.
We have shown a galaxy area observed with the SAURON at the continuum map.}
\label{mangal}
\end{figure*}

\section{DISCUSSION}

\subsection{Are There Young Stars in NGC 4143?}

\begin{figure*}[p]
\centerline{
\includegraphics[height=0.5\textwidth]{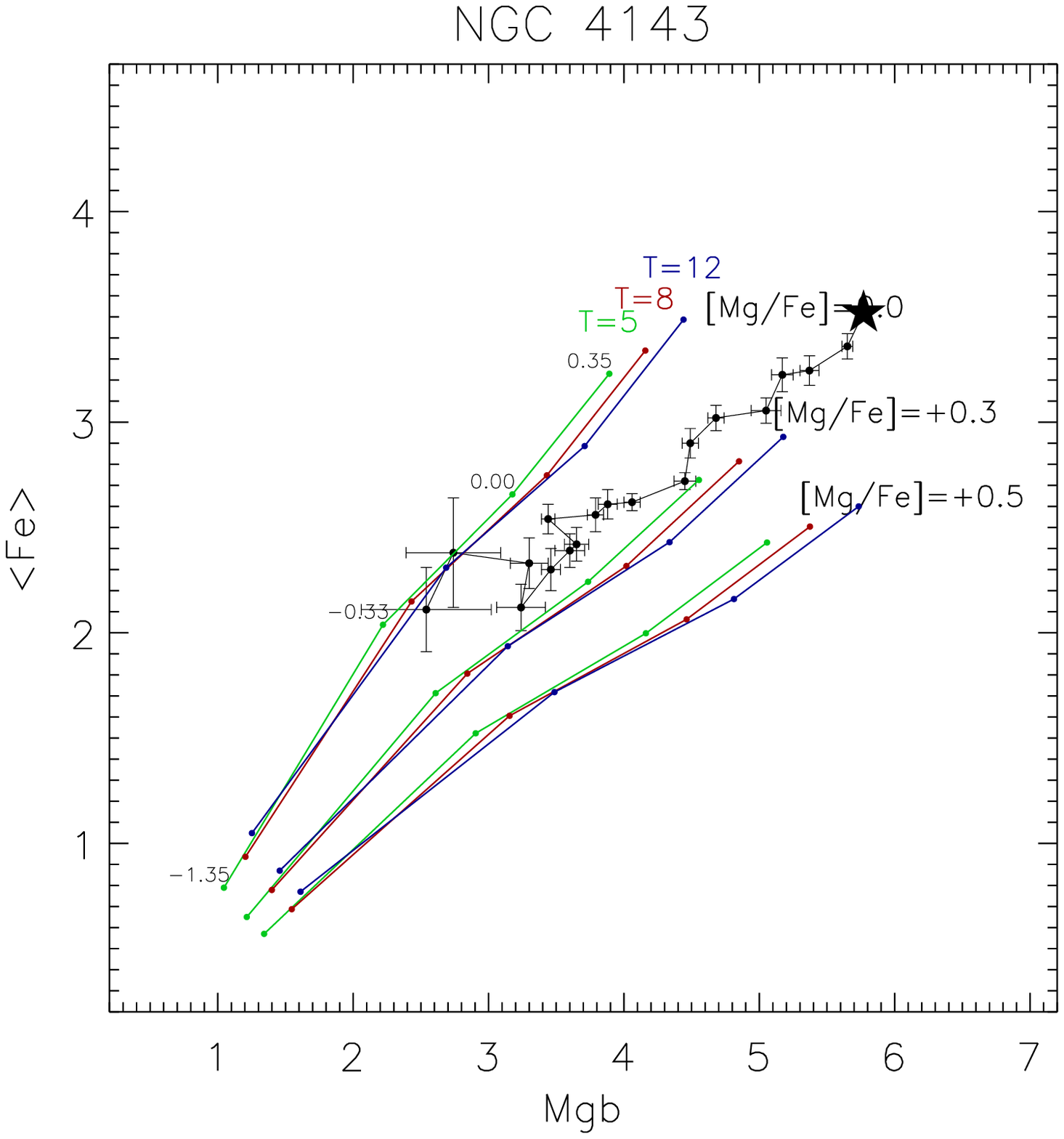}
\includegraphics[height=0.5\textwidth]{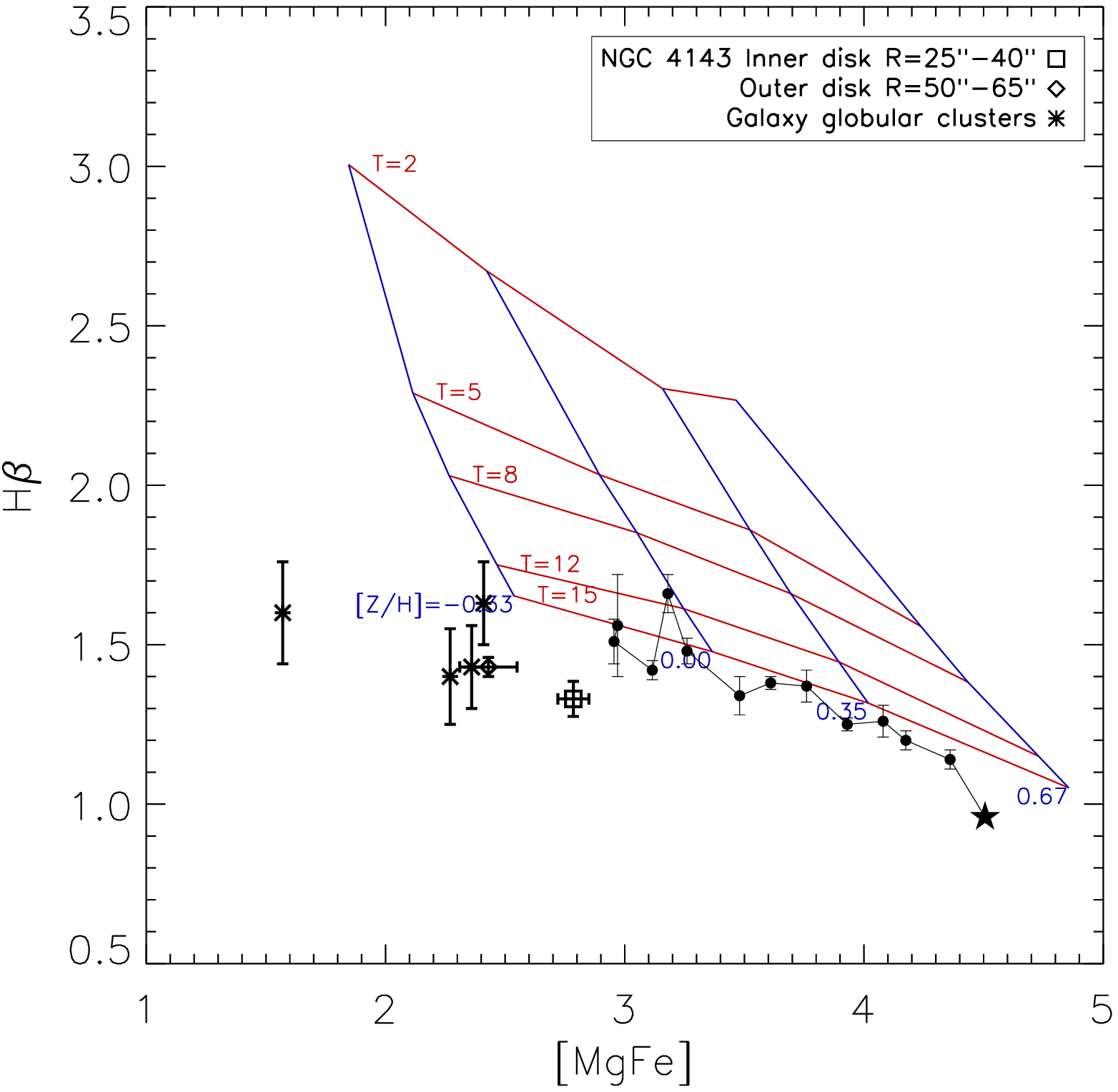}
}
\caption{Lick index-index diagrams for NGC~4143. 
The {\it left plot} represents Mgb vs iron index diagram which allows to estimate magnesium-to-iron ratio 
through the comparison of our measurements with the models by Thomas et al.(2003) for the different Mg/Fe ratios. 
By confronting the H$\beta$ Lick index versus a combined metallicity Lick index involving magnesium and iron lines 
({\it right plot}), we solve the metallicity-age degeneracy and determine both stellar-population parameters. 
Five different age sequences (red lines) are plotted as reference frame; 
the blue lines crossing the model age sequences mark the metallicities of $+0.67$, $+0.35$, 0.00, --0.33 from right to left.
A large black star corresponds to the central core, and then we go along the radius through the bulge;
the inner and outer parts of the stellar disk are plotted by different signs.
A few globular clusters from Beasley et al. (2004) belonging to the Galactic bulge are also plotted for the reference frame.} 
\label{inddiag}
\end{figure*}

We estimated the mean (luminosity-weighted) age of the stellar population along the radius of NGC~4143 by applying
the evolution model of 'simple stellar populations' -- SSP (Thomas et al. 2003), implying one short starburst, to
the Lick indices measured at various distances from the center. This model is exactly applicable for the
galactic nucleus and bulge, because our measurements of the Lick magnesium and iron indices show
that at the center of NGC~4143 the magnesium-to-iron abundance ratio is approximately twice the solar one (Fig. 6a); 
and this suggests a short duration of the main star formation epoch, less than 1 Gyr. In the galactic disk 
this ratio is close to the solar one, i.e., star formation there lasted longer than at least 2~Gyr.
However, the mean age of the stellar populations in NGC~4143 is homogeneously old everywhere, comparable to the
age of the Universe (Fig. 6b). The SSP models, into which one brief starburst to form the entire stellar population
of the galaxy/galactic region is built, have the mean stellar population age as their parameter, and this 'SSP-equivalent'
age is in fact the estimate of the time elapsed after the main starburst. If, however, we apply the SSP models to real 
systems, in which the star formation epoch had a long duration (for systems with a solar magnesium-to-iron ratio this
duration is at least 3~Gyr), then the age of the system estimated from these models is an age weighted with
the luminosity of stars. Since the young generations of stars, including massive stars, are always brighter
than the old ones, where the massive stars are already dead, this age estimate is shifted toward the starburst
finish time. In the case of the NGC~4143 disk, the combination of a solar magnesium-to-iron ratio and an old SSP 
age suggests that its formation ended at least 10 Gyr ago, while it most likely began approximately 13–--14 Gyr ago.

\begin{figure}[!h]
\includegraphics[scale=0.8]{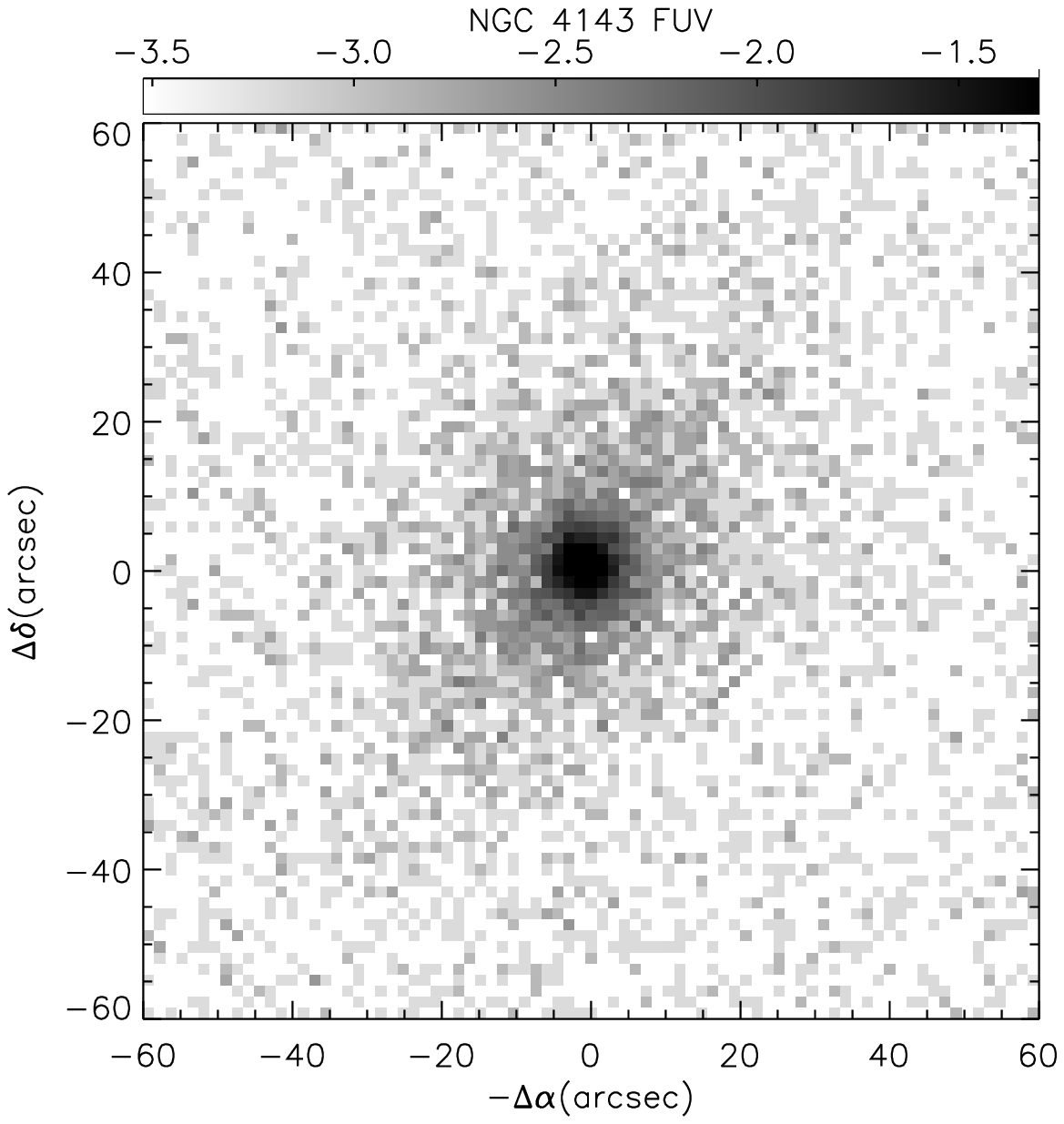}
\caption{
The map of NGC~4143 in the FUV-band of the GALEX space telescope ($\lambda_{mean}=1530$~\AA).}
\label{fuv}
\end{figure}

As we have noted above, our diagnostics based on the ratio of the fluxes of strong emission lines shows
that there is no noticeable current star formation in the galaxy, i.e., there are no HII regions in the
extended disk of NGC~4143. The integrated colors of the galaxy taken from the NED database, namely 
$NUV-r=5.59$ (the GALEX and SDSS/DR9 data) and $W2-W3=0.88$ (the WISE data), also unambiguously
assign NGC~4143 to the so-called passive galaxies (Kaviraj et al. 2007; Cluver et al. 2017).
However, the ultraviolet map of the galaxy that we retrieved from the open archive of the
data of the GALEX space telescope (Fig.~7) shows an extended disk with a radius of about $20^{\prime \prime}$ 
in the far ultraviolet (FUV, the effective wavelength is $\sim 1500$~\AA),
and an even more extended one, up to $50^{\prime \prime}$, in the near ultraviolet (NUV, the effective wavelength 
is $\sim 2300$~\AA). The exponential scalelength of this UV-disk, $14.9 \pm 1.0$ arcsec, coincides with what we obtained 
from the $r$-band SDSS image of the galaxy. In this case, we probably deal with an extreme manifestation of
the so-called UV upturn -- an ultraviolet excess in the spectra of galaxies with an old stellar population.
The combined ultraviolet–--optical color of the galaxy, $y = (NUV - u) - 1.7(u - g)$, is at the blue boundary
of the colors for a passive stellar population (Ali et al. 2019), i.e., shows a maximum ultraviolet excess.

\subsection{Does the Accretion Pattern Determine the Morphological Type of a Galaxy?}

Thus, the example of NGC~4143 confirms the trend reported by us previously (Sil'chenko et al. 2019):
no star formation proceeds in the gaseous disks inclined with respect to the stellar ones in lenticular
galaxies. In some sense, this galaxy is an extreme example: its extended gaseous disk is fully ionized by
a shock, while the neutral gas in NGC~4143 is not detected with a very low detection limit:
$\log M(\mbox{HI})< 6.80$ (Serra et al. 2012) and $\log M(\mbox{H2})<7.20$ (Young et al. 2011).
However, the stellar component associated with the accreted gas is probably also present
in the central region of the galaxy: the orientation of the central stellar disk estimated by the methods of
surface photometry is consistent with the kinematic estimates of the gaseous disk line-of-nodes parameters
within $15^{\prime \prime} - 20^{\prime \prime}$ from the center. This means that the minor merging of a gas-rich
satellite from an inclined orbit could be the source of gas accretion: a disrupted satellite is capable 
of providing both a gaseous component and a stellar one rotating in an inclined plane. 

Accretion of an external cold gas onto disk galaxies is currently believed to be the most important effect
that determines the entire course of galactic evolution (see, e.g., Combes 2015). If we turn to the general
scenario for the evolution of galaxies in the nearby Universe, then it seems from intuitive considerations
that galaxies of common mass, in an environment of similar density, should accrete an external cold gas with an
equal probability. Why do some of them (most) then form thin disks with current star formation, while others
(S0s) inherit the old stellar disks from early evolution epochs and now may have extended gaseous disks, but
do not replenish the stellar disk component by young stars? And still others (Es), as a rule, do not have
cold gaseous disks at all? This question is being actively discussed by astronomers at present. 
There is a point of view that these differences can be fully explained by the angular momentum of the
accreting gas (see, e.g., Peng and Renzini (2020) and references therein). Indeed, statistically, the gaseous
disks of S0 galaxies are more extended than those of spirals (Wang et al. 2016); may be the gas
infalling onto S0s does not reach the central disk regions, where it can become denser and ignite star
formation, due to its high angular momentum (Peng and Renzini 2020). In addition, apart from the angular
momentum of the infalling gas, we propose to consider yet another factor: the direction from which it
comes. Indeed, to explain the very strong evolution of the angular momentum of spiral galaxies during the last
8~Gyr, it should be assumed that an external gas is accreted onto them strictly in the plane of their stellar
disks, with the same spin orientation as the disk rotation spin (Renzini 2020). And what if the gas infalls
at an angle? Our studies (Sil'chenko et al. 2019) suggest that, firstly, it is often observed in S0 galaxies
and, secondly, in this case the infalling gas is heated by a shock and becomes incapable of star formation.
Thus, the difference in the morphological type of a disk galaxy can actually be a difference in the pattern
of gas accretion: both the direction of external-gas infall and how high is its orbital spin must play, as well
as of course the amount of mass inflowing. As regards the elliptical galaxies, the difference in the observed
X-ray flux from the hot gaseous galactic halos comes to mind here: in the field elliptical galaxies it is substantial
and has been easily measured by the Chandra and XMM-Newton X-ray telescopes (see, e.g., Mulchaey and Jeltema 2010), 
while in spiral galaxies of the same mass it is still barely detected. It may be that
the cold gas does not reach the elliptical galaxies, because it is heated during the inflow 
through the hot halo gas. It is then clear why no large-scale gaseous disks form usually around elliptical galaxies, 
even around those located in sparse environments.

\section{ACKNOWLEDGMENTS}

The paper is based on the observational data obtained with the 6-m telescope at the Special
Astrophysical Observatory of the Russian Academy of Sciences and with the 2.5-m telescope at the Caucasus
Mountain Observatory of the Sternberg Astronomical Institute of the Moscow State University. We
thank R.I. Uklein who performed the BTA observations. The whole work was supported by the MSU Development
Program (Leading Scientific School ''Physics of Stars, Relativistic Objects, and Galaxies''). The
observations at the BTA SAO RAS telescope are supported by the Ministry of Science and Higher
Education of the Russian Federation (including contract no. 05.619.21.0016, unique project identifier
RFMEFI61919X0016). In our analysis we used data from publicly accessible archives and databases:
the Lyon–--Meudon Extragalactic Database (LEDA) maintained by the LEDA team at the Lyon Observatory
CRAL (France) and the NASA/IPAC Extragalactic Database (NED) operated by the Jet Propulsion Laboratory
of the California Institute of Technology under contract with the National Aeronautics and Space Administration 
(USA), as well as some public SDSS, SDSS-II, and SDSS-III data (http://www.sdss3.org/) financed by the Alfred
P. Sloan Foundation, the participating institutes of the SDSS collaboration, the National Science 
Foundation, the US Department of Energy, the National Aeronautics and Space Administration (NASA), the
Japanese Monbukagakusho Foundation, the Max Planck Society, and the Financing Council for Higher
Education of England. We also invoked the data from the GALEX and WISE space telescopes for
our analysis. The NASA GALEX data were taken from the Mikulski Archive for Space Telescopes
(MAST). The data from the WISE space telescope used by us were taken from the NASA/IPAC Archive
operated by the Jet Propulsion Laboratory of the California Institute of Technology under contract with
the National Aeronautics and Space Administration. We also used the data retrieved from the Isaac
Newton Group Archive maintained as part of the CASU Astronomical Data Center at the Institute
of Astronomy of the Cambridge University (Great Britain).


\begin{thebibliography}{}

\bibitem[1]{scorpio2}
V. L. Afanasiev and A. V. Moiseev, Baltic Astronomy {\bf 20}, 363 (2011).

\bibitem[2]{sdss}
C. P. Ahn, R. Alexandroff, C. Allende Prieto, S. F. Anderson, T. Anderton, et al., ApJ Suppl. Ser. {\bf 203}, Aid.21 (2012). 

\bibitem[3]{ali19}
S. S. Ali, M. N. Bremer, S. Phillipps, and R. De Propris, MNRAS {\bf 487}, 3021 (2019).

\bibitem[4]{sauron}
R. Bacon, Y. Copin, G. Monnet, B. W. Miller, J. R. Allington-Smith, et al., MNRAS {\bf 326}, 23 (2001).

\bibitem[5]{glclust}
M. A. Beasley, J. P. Brodie, J. Strader, D. Forbes, et al., AJ {\bf 128}, 1623 (2004). 

\bibitem[6]{atlas3d}
M. Cappellari, E. Emsellem, D. Krajnovic, R. M. McDermid, N. Scott, et al., MNRAS {\bf 413}, 813 (2011).

\bibitem[7]{ewha}
R. Cid Fernandes, G. Stasi\'nska, M. S. Schlickmann, A. Mateus, N. Vale Asari, et al.,, MNRAS {\bf 403}, 1036 (2010).

\bibitem[8]{wise}
M. E. Cluver, T. H. Jarrett, D. A. Dale, J.-D. T. Smith, T. August, et al., ApJ {\bf 850}, Aid.68 (2017).

\bibitem[9]{combes15}
F. Combes, Highlights of Astronomy {\bf 16}, 366 (2015).

\bibitem[10]{atlaskin}
T. A. Davis, K. Alatalo, M. Sarzi, M. Bureau, L. M. Young, et al., MNRAS {\bf 417}, 882 (2011).

\bibitem[11]{erwin03}
P. Erwin and L. S. Sparke, ApJ Suppl. Ser. {\bf 146}, 299 (2003).

\bibitem[12]{erwin05}
P. Erwin, J. E. Beckman, and M. Pohlen, ApJ {\bf 626}, L81 (2005).

\bibitem[13]{erwin08}
P. Erwin, M. Pohlen, and J. E. Beckman, AJ {\bf 135}, 20 (2008).

\bibitem[14]{n4138}
K. P. Jore, A. H. Broeils, and M. P. Haynes, AJ {\bf 112}, 438 (1996).

\bibitem[15]{uma_kar}
I. D. Karachentsev, O. G. Nasonova O.G., and H. M. Courtois, MNRAS {\bf 429}, 2264 (2013).

\bibitem[16]{galex}
S. Kaviraj, K. Schawinski, J. E. G. Devriendt, I. Ferreras, S. Khochfar, et al., ApJ Suppl. Ser. {\bf 173}, 619 (2007).

\bibitem[17]{kewley}
L. J. Kewley and S. L. Ellison, ApJ {\bf 681}, 1183 (2008).

\bibitem[18]{kgo}
V. Kornilov, B. Safonov, M. Kornilov, N. Shatsky, O. Voziakova, S. Potanin, et al., PASP {\bf 126}, 482 (2014).

\bibitem[19]{lauri10}
E. Laurikainen, H. Salo, R. Buta, J. H. Knapen, and S. Comer\'on, MNRAS {\bf 405}, 1089 (2010). 

\bibitem[20]{lauri11}
E. Laurikainen, H. Salo, R. Buta, and J. H. Knapen, MNRAS {\bf 418}, 1452 (2011).

\bibitem[21]{lovelace}
R. V. E. Lovelace, K. P. Jore, and M. P. Haynes, ApJ {\bf 475}, 83 (1997).

\bibitem[22]{moisav04}
A. V. Moiseev, J. R. Vald\'es, and V. H. Chavushyan, A\&A {\bf 421}, 433 (2004).

\bibitem[23]{mangal}
A.  Moiseev, A. Perepelitsyn, and D. Oparin, Experimental Astronomy, submitted (2020).

\bibitem[24]{xray}
J. S. Mulchaey and T. E. Jeltema, ApJ {\bf 715}, L1 (2010).

\bibitem[25]{pak14}
M. Pak, S.-C. Rey, T. Lisker, Y. Lee, et al., MNRAS {\bf 445}, 630 (2014).

\bibitem[26]{peng_renzini}
Y.-j. Peng and A. Renzini, MNRAS {\bf 491}, L51 (2020).

\bibitem[27]{pogge_esk}
R. W. Pogge and P. B. Eskridge, AJ {\bf 106}, 1405 (1993).

\bibitem[28]{renzini20}
A. Renzini, MNRAS {\bf 495}, L42 (2020).

\bibitem[29]{sage_welch}
L. J. Sage and G. A. Welch, ApJ {\bf 644}, 850 (2006).

\bibitem[30]{atlas_h1}
P. Serra, T. Oosterloo, R. Morganti, K. Alatalo, L. Blitz, et al., MNRAS {\bf 422}, 1835 (2012).

\bibitem[31]{silsau}
O.K. Sil'chenko, Astron. Lett. {\bf31}, 227 (2005).  

\bibitem[32]{sil_fp}
O. K. Sil'chenko, A. V. Moiseev, and O. V. Egorov, ApJ Suppl. Ser. {\bf 244}, Aid. 6 (2019). 

\bibitem[33]{thomod}
D. Thomas, C. Maraston, R. Bender, MNRAS {\bf 339}, 897 (2003)

\bibitem[34]{tully_uma}
R. B. Tully, M. A. W. Verheijen, M. J. Pierce, J.-S. Huang, and R. J. Wainscoat, AJ {\bf 112}, 2471 (1996).

\bibitem[35]{barsim}
P. Vauterin and H. Dejonghe, MNRAS {\bf 286}, 812 (1997). 

\bibitem[36]{uma_h1}
M. A. W. Verheijen and R. Sancisi, A\&A {\bf 370}, 765 (2001).

\bibitem[37]{wakamatsu}
K.-I. Wakamatsu, AJ {\bf 105}, 1745 (1993).

\bibitem[38]{bluedisk}
J. Wang, B. S. Koribalski, P. Serra, T. van der Hulst, S. Roychowdhury, et al., MNRAS {\bf 460}, 2143 (2016).

\bibitem[39]{welch_sage}
G. A. Welch and L. J. Sage, ApJ {\bf 584}, 260 (2003).

\bibitem[40]{welch10}
G. A. Welch, L. J. Sage, and L. M. Young, ApJ {\bf 725}, 100 (2010).

\bibitem[41]{atlas_molgas}
L. M. Young, M. Bureau, T. A. Davis, F. Combes, R. M. McDermid, et al., MNRAS {\bf 414}, 940 (2011).

\end{thebibliography}
\end{document}